\documentclass[11pt]{article}

\usepackage[margin=0.75in]{geometry}  % adjust margins similar to informs3.cls
\usepackage{setspace}                 % for spacing control
\usepackage{natbib}                   % citations
\usepackage{graphicx}                 % images
\usepackage{subcaption}              % subfigures
\usepackage{caption}
\usepackage{threeparttable}
\usepackage{booktabs}
\usepackage{amsmath, amssymb}        % math packages
\usepackage{palatino, mathpazo}      % fonts (palatino)
\usepackage{todonotes}     % todo notes

\usepackage[hyphens]{url}     % enables line breaks in URLs

\bibpunct[, ]{(}{)}{,}{a}{}{,}
\title{Zara: An LLM-based Candidate Interview Feedback System}
\author{
  Nima Yazdani \\ University of Southern California, micro1 \\ \texttt{nima@micro1.ai} \\\\
  Aruj Mahajan \\ micro1 \\ \texttt{aruj@micro1.ai} \\\\
  Ali Ansari \\ Stanford University, micro1 \\ \texttt{ali@micro1.ai}\\
}
\date{}

\begin{document}
\maketitle

\begin{abstract}
This paper introduces Zara, an AI-driven recruitment support system developed by micro1, as a practical case study illustrating how large language models (LLMs) can enhance the candidate experience through personalized, scalable interview support. Traditionally, recruiters have struggled to deliver individualized candidate feedback due to logistical and legal constraints, resulting in widespread candidate dissatisfaction. Leveraging OpenAI's GPT-4o, Zara addresses these limitations by dynamically generating personalized practice interviews, conducting conversational AI-driven assessments, autonomously delivering structured and actionable feedback, and efficiently answering candidate inquiries using a Retrieval-Augmented Generation (RAG) system. To promote transparency, we have open-sourced the approach Zara uses to generate candidate feedback.
\end{abstract}

\section{Introduction} \label{intro}
Recruitment processes have long been criticized for leaving candidates without actionable feedback. Over half of candidates receive no feedback after screening or interviews, and nearly 70\% remain uninformed even after final-stage rejections; when feedback is provided, more than three-quarters find it too generic to be useful \citep{Mead2020}. This lack of communication frustrates candidates, negatively affects psychological well-being, and damages employer reputations, as candidates tend to share negative experiences widely \citep{Mead2020, wood2023ghosting}.
Recent advances in large language models (LLMs) offer an opportunity to address these issues by delivering personalized feedback at scale. Models like GPT-4o generate fluent, context-aware responses and have effectively handled personalized interactions in domains such as customer service \citep{limna2023role} and HR communications \citep{mujtaba2024fairness}. By leveraging these capabilities, LLMs can potentially overcome logistical barriers to providing structured, individualized candidate feedback.

\begin{figure}[htbp] \centering \includegraphics[width=0.6\textwidth]{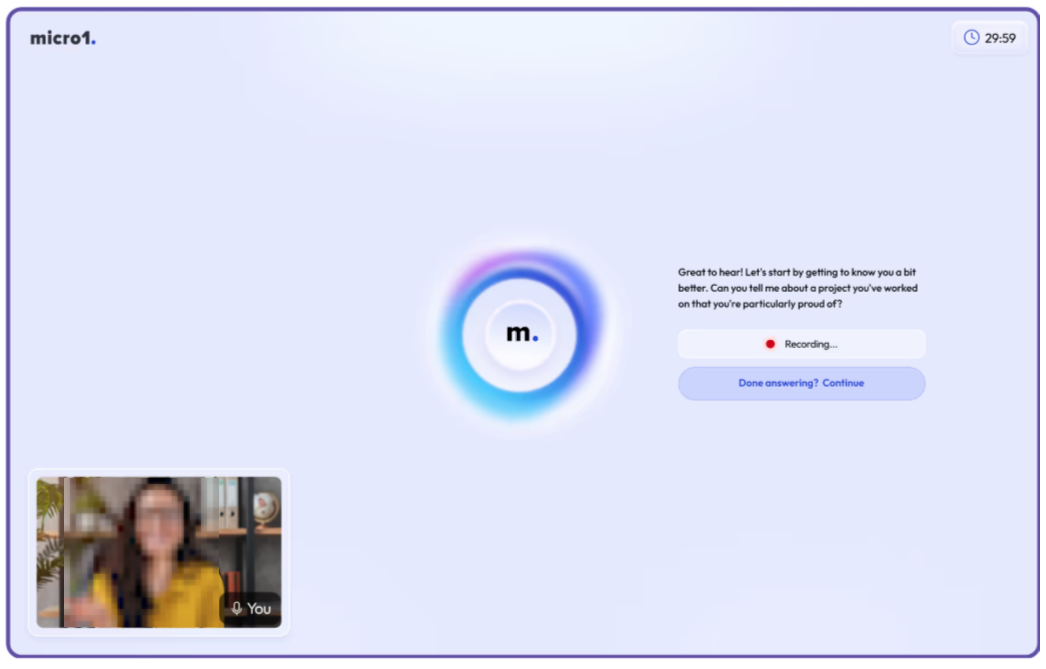} \caption{The Zara AI-led interview interface integrated into micro1’s recruitment platform.} \label{fig:InterviewScreenshot} \end{figure}

\begin{figure}[htbp] \centering \includegraphics[width=0.6\textwidth]{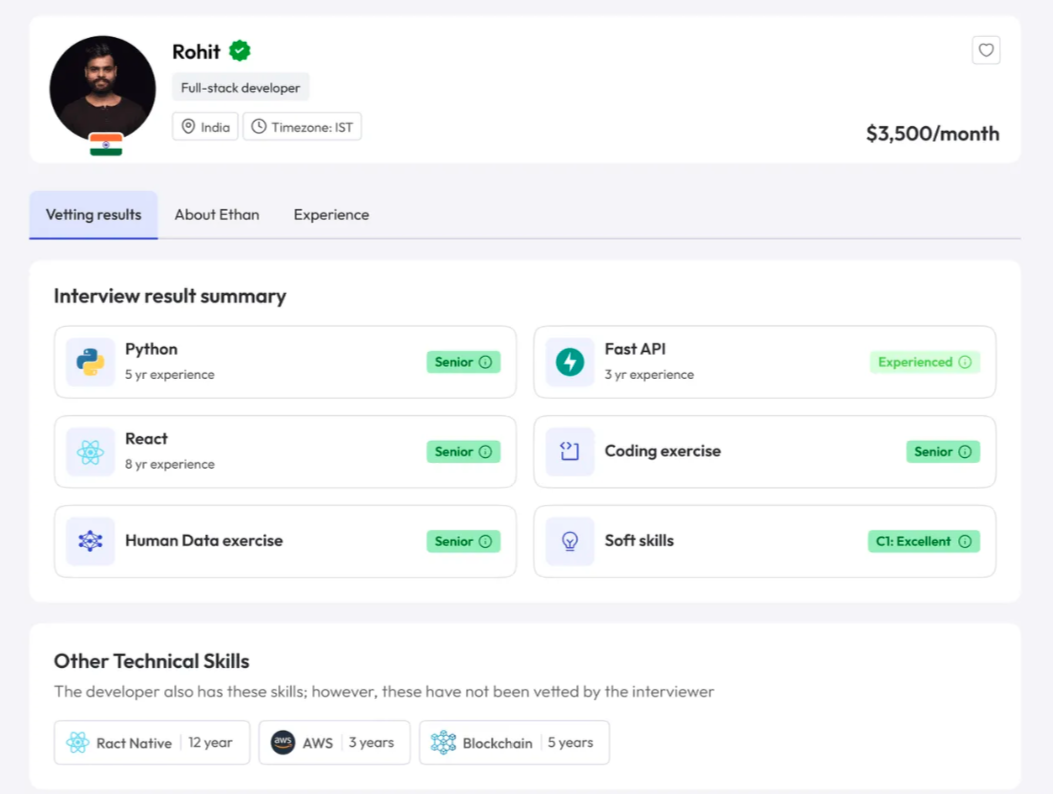} \caption{An example profile report generated by Zara and added to micro1's talent pool.} \label{fig:report} \end{figure}

This paper presents \textit{Zara}, an AI-driven recruiter primarily powered by OpenAI's GPT-4o, supplemented by additional LLMs, fine-tuned models, and a Retrieval-Augmented Generation (RAG) framework. Integrated into micro1’s hiring platform, Zara provides personalized practice interviews, conducts conversational AI-led interviews tailored specifically to evaluate a candidate's technical and interpersonal skills for the targeted job role, autonomously generates structured post-interview evaluations based directly on the candidate's performance during the AI-led interview, and answers candidate's questions throughout the recruitment process. Employers use Zara to efficiently identify promising candidates, whose AI-generated assessments are subsequently reviewed by human recruiters. An example of the Zara interview interface can be seen in Figure~\ref{fig:InterviewScreenshot}. Candidates who excel in their virtual interview with Zara, progress to subsequent human interview rounds, and their profiles, including detailed assessments of technical and soft skills, are added to micro1’s talent pool, facilitating consideration for additional opportunities from future employers (see Figure~\ref{fig:report} for a sample AI-generated report). Those applicants who do not meet these criteria have the option to request detailed, actionable feedback via email, directly derived from their interview performance.


 The primary objective of this paper is to open-source Zara's candidate support framework, with particular emphasis on its innovative personalized feedback and candidate inquiry management capabilities, and to share insights gained from deploying this LLM-driven system in practice. By presenting Zara as an applied use case rather than a novel algorithm, we illustrate how existing LLM capabilities significantly improve candidate experiences at scale. Subsequent sections explore historical barriers to recruitment feedback, review related applications of LLM technology, and detail Zara’s architecture, implementation, and results.

\section{Background}

\subsection{Lack of Feedback in Recruitment}

Companies traditionally provide minimal or no feedback due to high candidate volumes, limited recruiter time, and perceived legal risks associated with candid critiques potentially leading to discrimination disputes \citep{Odoro2024}. Many organizations default to minimal communication or complete silence, known as “ghosting,” leaving applicants uncertain about their performance and reducing their sense of fairness and preparedness for future opportunities \citep{wood2023ghosting}. Moreover, companies miss opportunities to positively engage candidates, who would be more likely to reapply or recommend the employer if they received constructive feedback \citep{Mead2020}. This persistent feedback gap thus represents both a human and business challenge, motivating scalable solutions.

\subsection{LLMs in Recruiting and Service Domains}

Recruiters have increasingly adopted AI and LLM tools to enhance efficiency and consistency across hiring stages. LLMs have successfully automated tasks such as composing job advertisements, refining job descriptions, and conducting standardized screening interviews through chatbots \citep{mujtaba2024fairness}. However, these tools typically have not provided personalized interview feedback. Additionally, academic research highlights potential biases in AI-generated interview content, underscoring the importance of careful implementation to ensure fairness and accuracy \citep{kong2024gender}.

In customer service and related fields, LLM-driven conversational agents successfully handle personalized interactions at scale, consistently improving service efficiency and user satisfaction \citep{limna2023role}. These experiences provide a useful blueprint for applying similar solutions in recruitment contexts. Effective implementation requires maintaining respectful, constructive, and human-centric communications aligned with HR best practices to augment recruiting capabilities without losing the personal touch \citep{sigala2024chatgpt}.

The recognized need for improved candidate feedback combined with mature LLM technology presents an ideal opportunity for innovation in recruitment processes, forming the foundation for Zara’s development, deployment, and expansion.

\section{Methodology}

\subsection{Design Overview}
This study adopts a comprehensive case study approach to examine the use of LLMs within the recruitment process. Zara leverages existing advanced LLM technologies, notably GPT-4o, rather than introducing novel algorithms.

Candidates interact with Zara through four integrated phases
%, shown in Figure~\ref{fig:RecruitmentLifecycle}
: (1) personalized interview preparation, (2) the AI-led interview, (3) structured post-interview feedback, and (4) automated candidate query resolution. This sequence reflects the typical candidate experience. In this section, we briefly touch on phase (1) and (2) but focus specifically on phases (3) and (4), as these form the core of Zara's candidate support capabilities.

% \begin{figure}[htbp]
% \centering
% \includegraphics[width=0.6\textwidth]{figures/RecruitmentLifecycle.png}
% \caption{The four phases of Zara's integration into recruitment: interview preparation, AI-led interview, personalized post-interview feedback, and automated candidate query handling.}
% \label{fig:RecruitmentLifecycle}
% \end{figure}

\begin{figure}[htbp]
\centering
\includegraphics[width=0.7\textwidth]{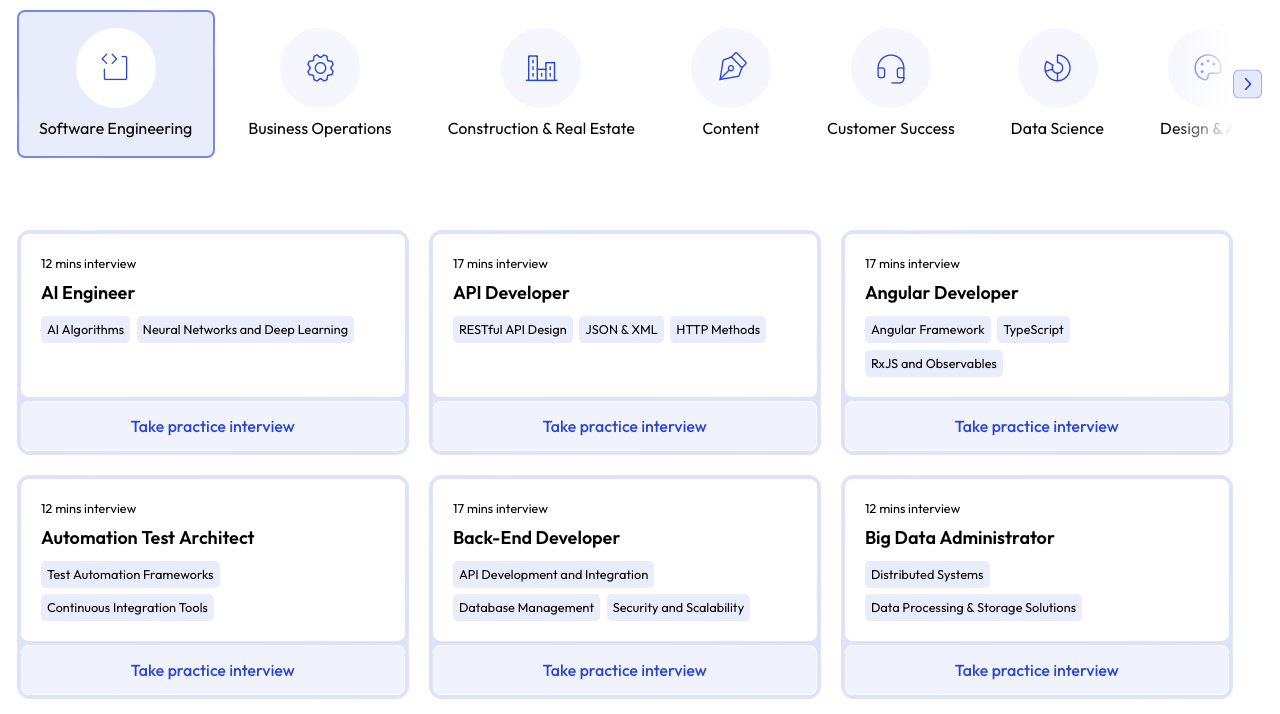}
\caption{Sample practice interviews generated per role and skill set.}
\label{fig:practice}
\end{figure}

\textbf{Personalized Interview Preparation:} Candidates first engage with Zara through customized practice interviews tailored to their targeted roles and skill sets, as illustrated in Figure~\ref{fig:practice}. These mock interview scenarios, help candidates familiarize themselves with the anticipated interview questions.

\textbf{AI-led Interview:}
Next, candidates undergo an AI-led virtual interview (Figure~\ref{fig:InterviewScreenshot}). Zara serves as an autonomous interviewer engaging candidates in structured yet natural conversations, evaluating their technical competencies and conversational abilities. The model dynamically adjusts questions based on candidates' responses, providing an objective and scalable assessment while also answering any clarification questions the candidate has mid-interview.

\textbf{Structured Post-Interview Feedback:} Following the interview, unsuccessful candidates, as defined in Section \ref{intro}, can request to receive structured feedback. Zara employs a chain-of-thought prompting approach \citep{wei2022chain}, first analyzing interview transcripts and assessments to form an overall evaluation. It identifies two to three specific strengths and two to three improvement areas, explicitly excluding soft skills and communication to reduce subjectivity. Zara then performs a reflective review, refining the feedback language to be constructive, supportive, and encouraging. Few-shot prompting techniques \citep{lee2024few} provide exemplar outputs to maintain consistent feedback formatting.

\textbf{Candidate Query Resolution:} Finally, Zara manages candidate inquiries using a Retrieval-Augmented Generation (RAG) approach. Candidate questions are vectorized into embeddings and matched against a precompiled FAQ database encoded similarly. A similarity threshold ensures only high-confidence matches trigger automated responses generated by GPT-4o. This automation significantly reduces human workload and maintains consistent, timely candidate support. This step is particularly important, as the volume of inquiries far exceeds what a team of support representatives can comprehensively answer.

\section{Implementation of Candidate Feedback and Query Resolution}

Zara uses GPT-4o as its primary LLM for all candidate interactions, including generating practice scenarios, performing AI-led interviews, producing structured feedback, and resolving candidate inquiries.

\subsection{Interview Feedback Generation}

Feedback generation leverages chain-of-thought prompting. An overview of this is presented in Figure~\ref{fig:COT_Feedback}. Zara first reviews interview transcripts to identify two to three specific candidate strengths and areas for improvement, deliberately excluding soft skills to maintain focus on tangible, actionable competencies. Feedback undergoes a reflective review step, ensuring it remains constructive, supportive, and positively phrased. Few-shot prompting with example outputs ensures consistency in tone and format, while guardrails systematically prevent overly negative language. The explicit prompts used for feedback generation are detailed below:

\begin{figure}[htbp]
\centering
\includegraphics[width=0.5\textwidth]{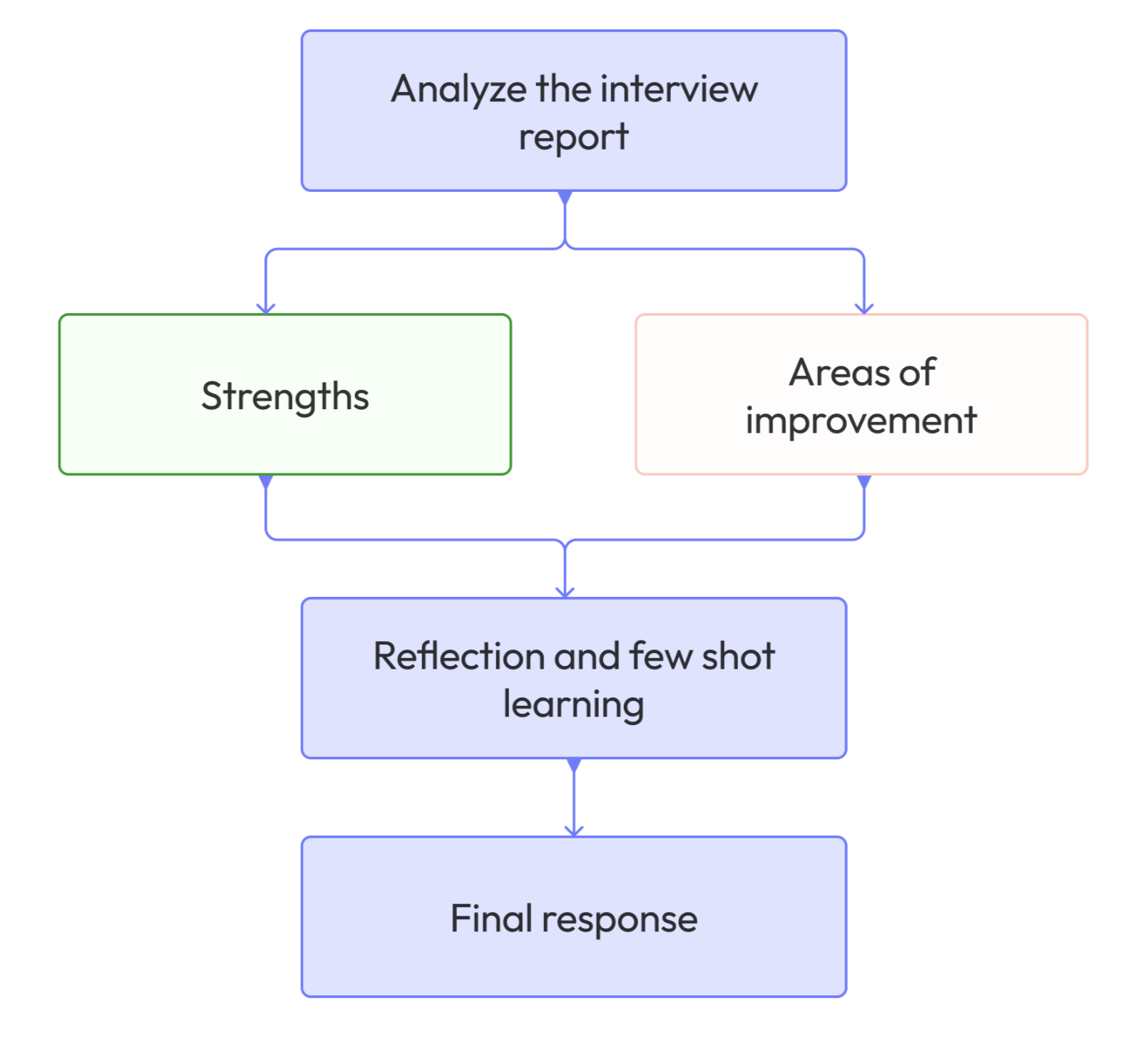}
\caption{Chain-of-Thought prompting technique for Zara interview feedback generation.}
\label{fig:COT_Feedback}
\end{figure}

\begin{quote}
\textbf{System Prompt:} \textit{"You are an experienced HR professional tasked with providing constructive feedback to job candidates based on their interview reports. Your goal is to create feedback that is specific, constructive, supportive, sugar-coated, and focuses on the candidate's strengths and areas for improvement. This feedback will help candidates understand their performance and guide their future development."}
\end{quote}

\begin{quote}
\textbf{User Prompt:} ``Here is the interview report you need to analyze:

\texttt{<interview\_report>\{INTERVIEW\_REPORT\}</interview\_report>}

Please follow these steps to generate the feedback:

\begin{enumerate}
    \item Carefully read and analyze the interview report.
    \item Identify 2-3 key strengths of the candidate.
    \item Identify 2-3 areas where the candidate could improve and refrain from giving any feedback on soft skills and communication.
    \item For each strength and area of improvement, formulate feedback in 1-2 sentences that is:
    \begin{itemize}
        \item Specific and backed by examples from the report.
        \item Constructive and forward-looking.
        \item Encouraging and supportive with sugar-coated language.
        \item Never assume they don't have experience; always assume they have experience and suggest improvements in past-tense phrasing (e.g., "could have explained better").
    \end{itemize}
\end{enumerate}

Structure your feedback in JSON format with two main keys: "strengths" and "areas\_for\_improvement," each containing 2-4 items. Each item includes:
\begin{itemize}
    \item A "title" key summarizing the feedback point.
    \item A "detail" key offering constructive advice.
\end{itemize}

\textit{Maintain a positive tone, avoid specific ratings such as senior or junior, and refer to the candidate directly as 'you' or 'your.' Do not return any additional information other than the requested JSON."}
\end{quote}

Candidate interactions originate within micro1’s interfaces, triggering API calls to Zara’s backend services. Role-specific data and candidate skill profiles flow into GPT-4o modules for generating customized scenarios and conducting AI-led interviews. Interview data then securely moves via internal APIs into Zara’s feedback generation pipeline, where they are processed. Structured feedback is securely returned to micro1 and delivered to candidates by email.

\subsection{Candidate Query Handling}
\begin{figure}[htbp]
\centering
\includegraphics[width=0.6\textwidth]{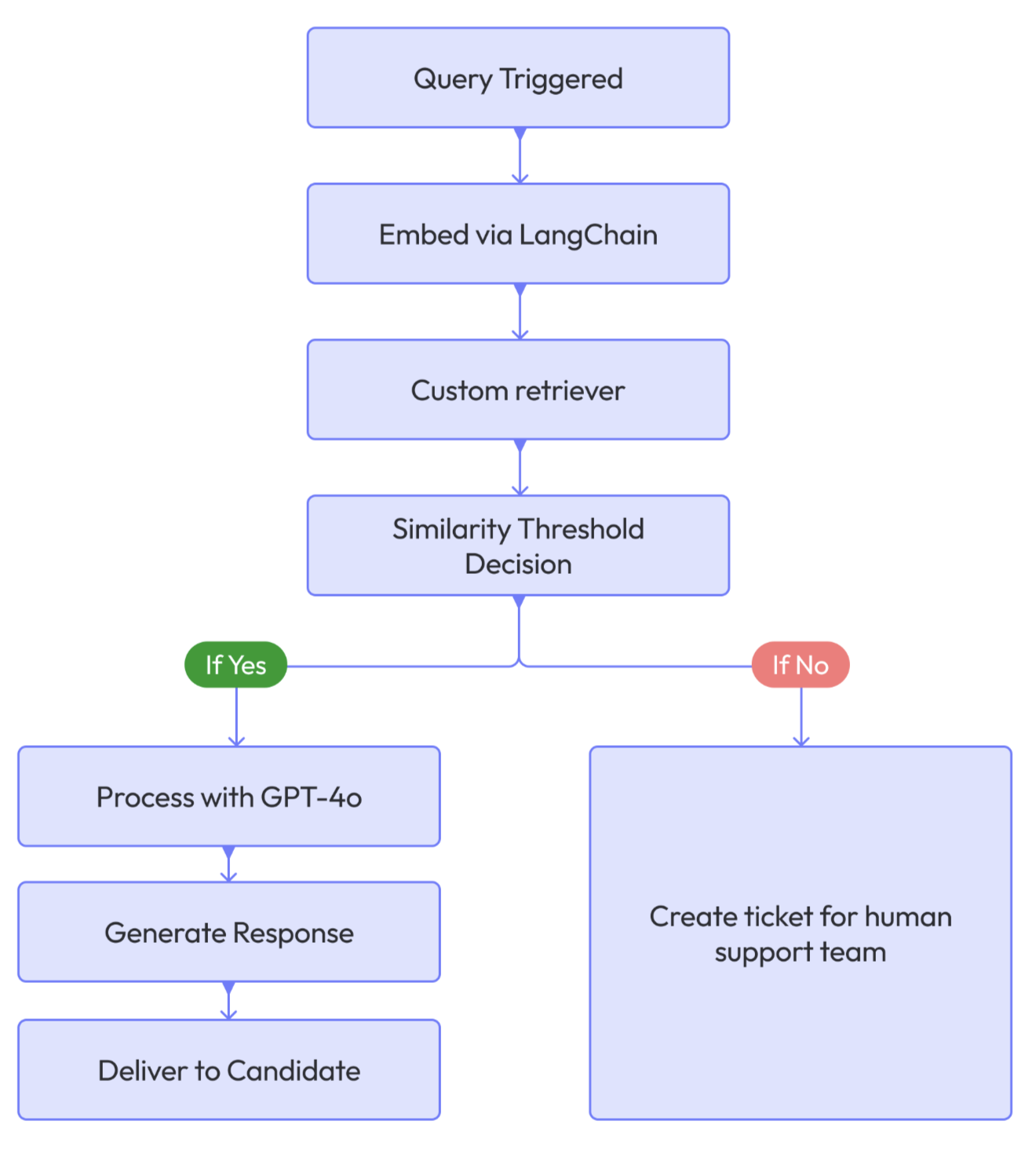}
\caption{Chain-of-Thought prompting technique for Zara candidate query response generation.}
\label{fig:COT_Rag}
\end{figure}

Candidate queries trigger Zara’s Retrieval-Augmented Generation (RAG) workflow powered by LangChain. Queries are converted into vector embeddings and matched via cosine similarity exclusively against preprocessed question embeddings stored in ChromaDB. By embedding only questions and excluding answer texts, Zara reduces redundancy and improves response relevance. A similarity threshold ensures that only high-confidence matches provide GPT-4o with precise context, enabling Zara to quickly generate coherent, accurate responses delivered directly to candidates. An overview of this RAG-based workflow is presented in Figure~\ref{fig:COT_Rag}. Zara’s modular API-centric architecture ensures that the handling of candidate queries is scalable, can be efficiently maintained, and easily integrated into the micro1 recruitment ecosystem.

\section{Results and Analysis}

\subsection{Usage Metrics and Candidate Engagement}

The implementation of Zara at micro1 was evaluated based on data collected from the platform's usage over a 3 day period. There were 4820 "unsuccessful" interviews conducted, of which \textbf{~10.7\%} requested detailed feedback. Furthermore, Zara's automated query handling system successfully resolves \textbf{~75\%} of candidate inquiries without requiring human intervention.

Candidate satisfaction with the AI-led interview experience was measured through Net Promoter Score (NPS). Based on 400 candidate ratings, Zara-enabled AI interviews received an average NPS rating of 4.37 out of 5.

\begin{figure}[htbp]
    \centering
    \begin{subfigure}[b]{0.48\textwidth}
        \centering
        \includegraphics[width=\textwidth]{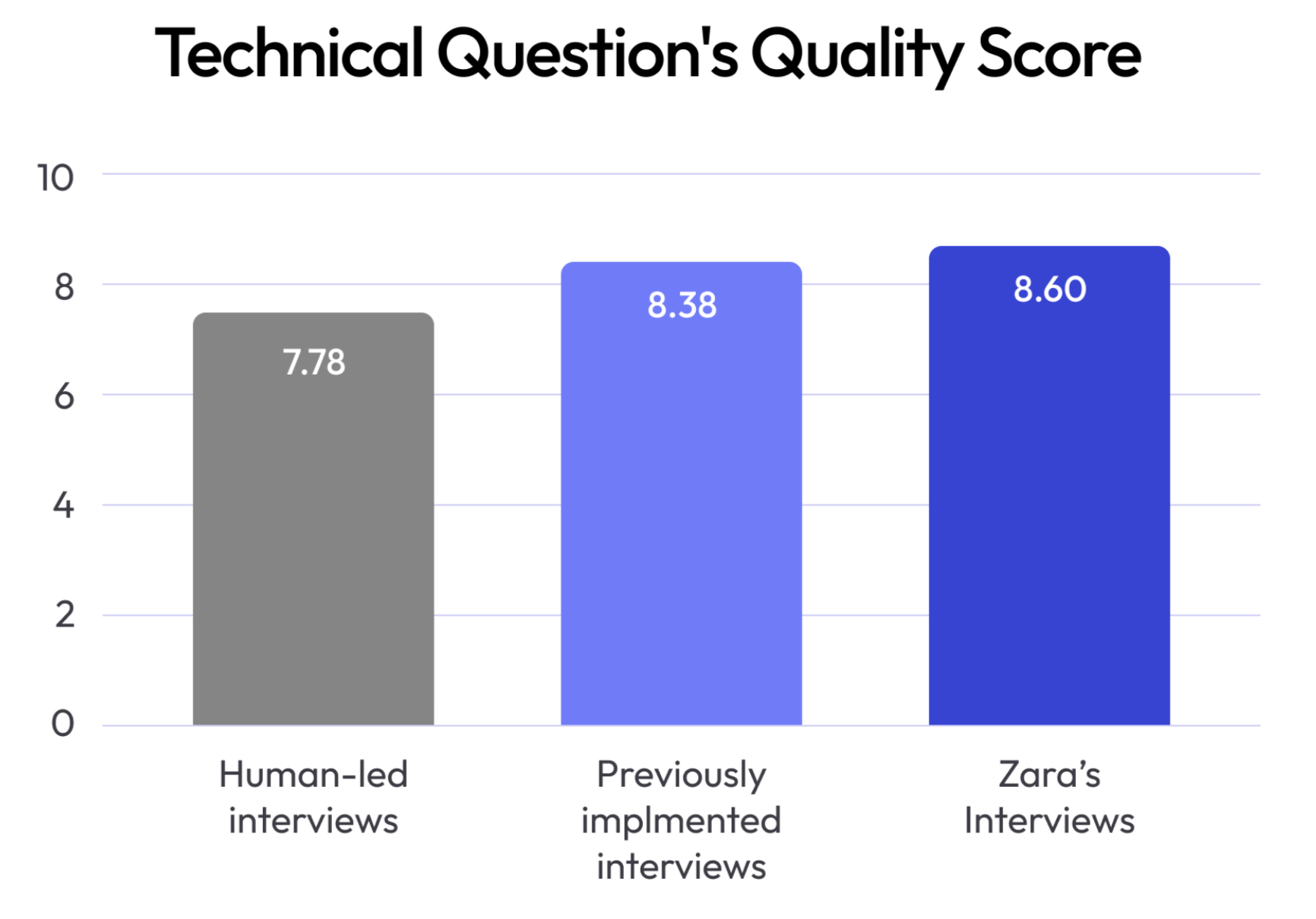}
        \caption{Comparison of Technical Question Quality Scores between Human-led Interviews, previously implemented interview structure, and Zara’s interviews.}
        \label{fig:technical_scores}
    \end{subfigure}
    \hfill
    \begin{subfigure}[b]{0.48\textwidth}
        \centering
        \includegraphics[width=\textwidth]{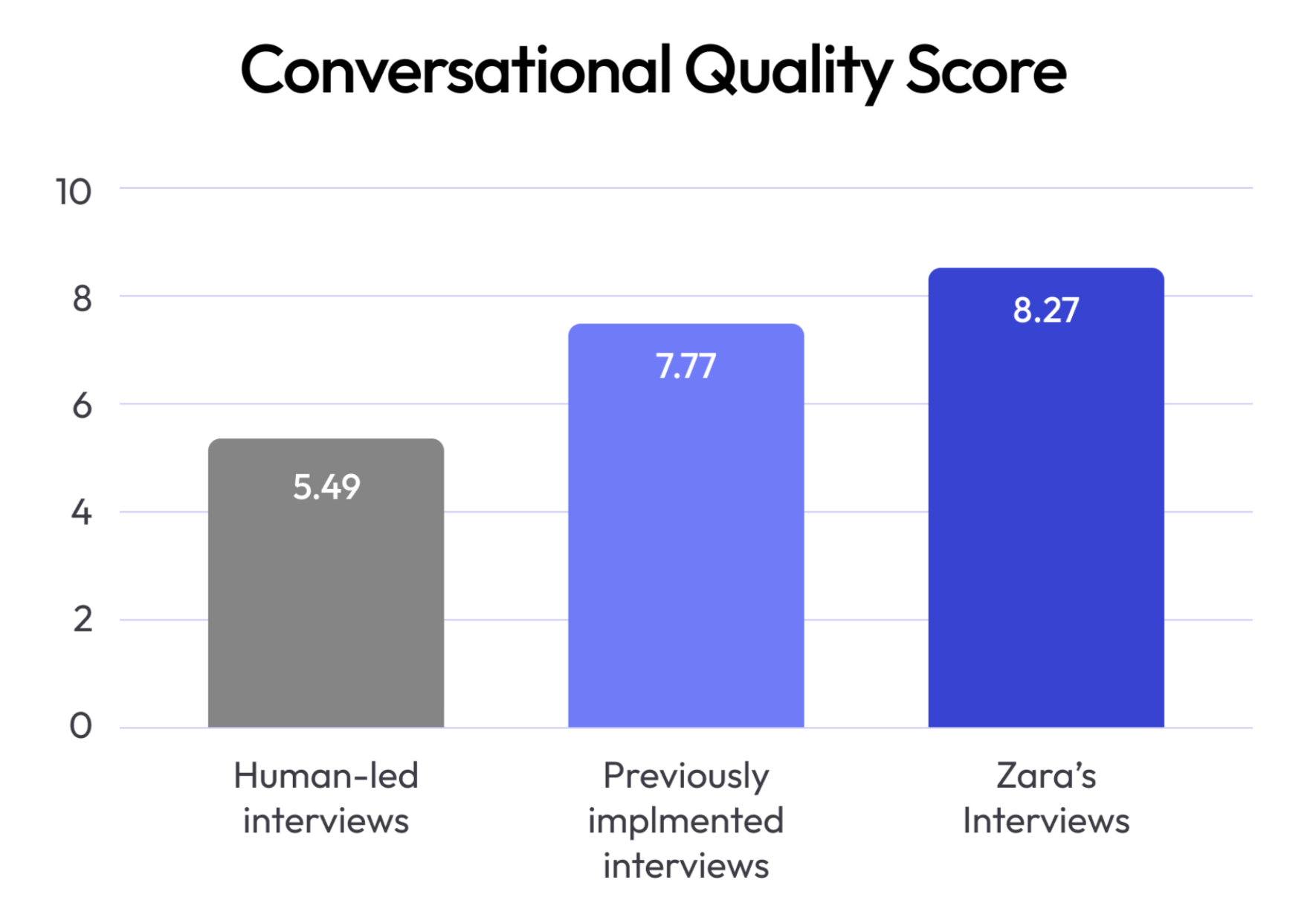}
        \caption{Comparison of Conversational Dynamics Quality Scores between Human-led Interviews, previously implemented interview structure, and Zara’s interviews.}
        \label{fig:conversational_scores}
    \end{subfigure}
    \caption{Comparative evaluation of interview quality metrics before and after Zara's implementation.}
    \label{fig:comparison_figures}
\end{figure}

Internal evaluations demonstrated clear improvements in both technical question quality and conversational dynamics following Zara’s integration, compared to micro1's previous GPT-4o-powered but non-fully conversational system. Technical question quality ratings increased from 8.38 in the prior format to 8.60, partly due to Zara’s capability to dynamically respond to candidate requests for clarification, additional context, and explanations, enabling more precise and effective questioning. Similarly, conversational dynamics ratings improved significantly from 7.77 to 8.27, reflecting Zara’s interactive approach that empowers candidates to engage actively and seek clarity throughout the interview. Figures~\ref{fig:technical_scores} and~\ref{fig:conversational_scores} illustrate these enhancements compared to traditional human-led interviews as well (7.78 for technical question quality and 5.49 for conversational dynamics).

\subsection{Example Detailed Candidate Feedback Reports}

\begin{figure}[htbp]
    \centering
    \includegraphics[width=0.9\textwidth]{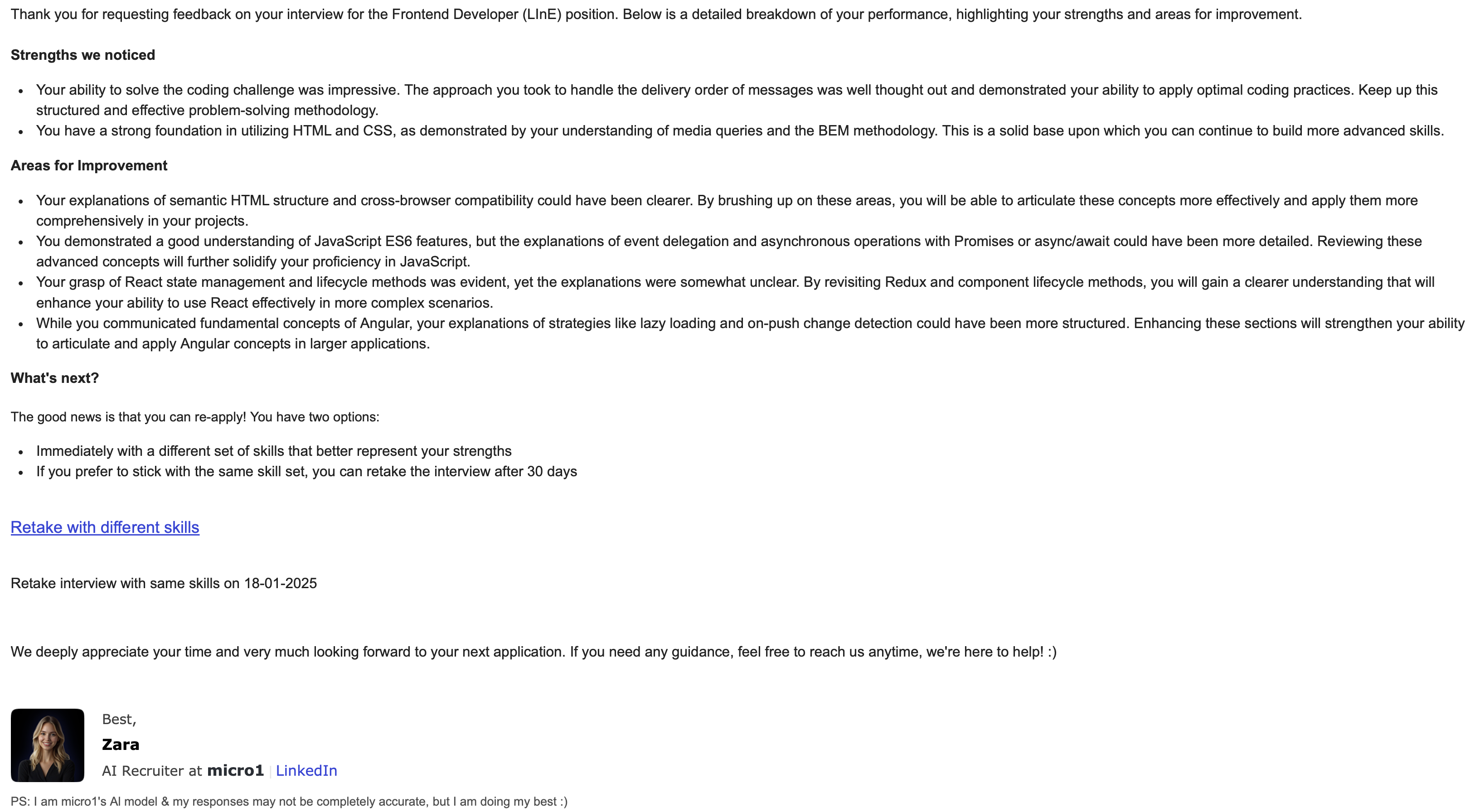}
    \caption{Example Feedback Report for a Frontend Developer candidate provided by Zara.}
    \label{fig:feedback_Software}
\end{figure}

\begin{figure}[htbp]
    \centering
    \includegraphics[width=0.9\textwidth]{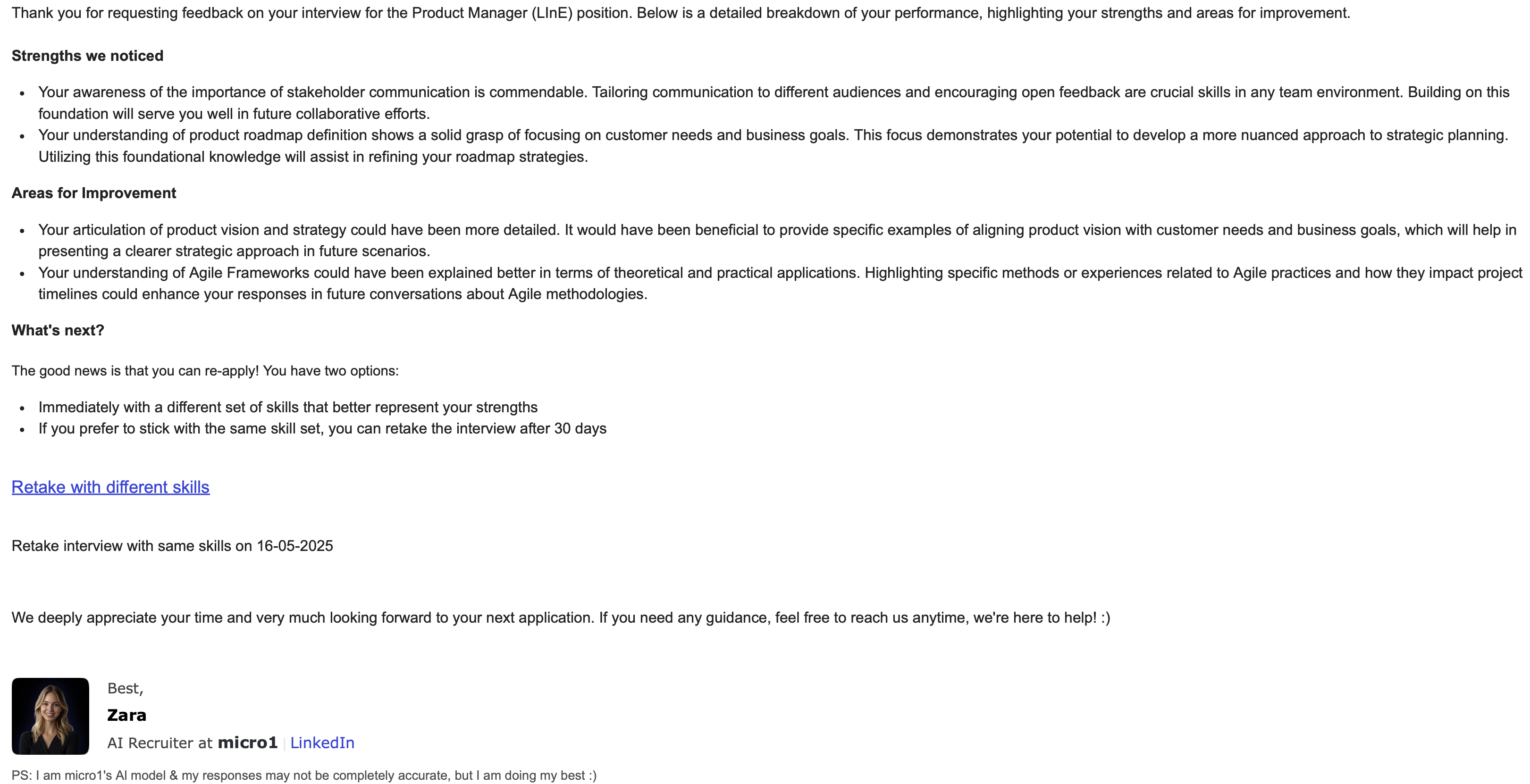}
    \caption{Example Feedback Report for a Product Manager candidate provided by Zara.}
    \label{fig:feedback_PM}
\end{figure}

Figures~\ref{fig:feedback_Software} and \ref{fig:feedback_PM} illustrate real examples of the detailed, structured feedback reports Zara provides to candidates following AI-led interviews. Each feedback report explicitly outlines candidate strengths and areas for improvement, consistently providing clear, actionable guidance directly related to the candidate’s interview performance. For instance, the feedback provided to the Frontend Developer candidate (Figure~\ref{fig:feedback_Software}) highlights the candidate's strong coding challenge performance and foundational skills in HTML and CSS, while recommending clearer explanations of semantic HTML structure, event delegation in JavaScript, React lifecycle methods, and Angular concepts. The feedback for the Product Manager candidate (Figure~\ref{fig:feedback_PM}) acknowledges their strengths in stakeholder communication and product roadmap development, while advising improved detail in articulating product vision and clearer explanations of theoretical and practical applications of Agile frameworks.

The consistent and structured nature of these feedback reports across various roles demonstrates Zara's capability to deliver clear, actionable, and role-specific feedback, significantly exceeding current recruitment standards and effectively addressing a longstanding challenge in traditional recruitment processes.

% \subsection{Interpretation of Findings}

% Quantitative results clearly demonstrate measurable improvements in candidate experience following Zara’s integration, particularly in technical question quality and conversational dynamics. These improvements directly correlate with the observed high candidate satisfaction scores (NPS: 4.37/5), suggesting that candidates strongly value a dynamic, conversationally fluid interview experience that allow for the candidate to ask for context, rephrasing, or clarification.

% Comparing metrics before and after Zara’s deployment indicates clear positive impact. Technical question quality scores improved modestly, while conversational dynamics ratings improved substantially, indicating candidates appreciate Zara’s approach. These improvements likely contributed directly to the significant increase in candidate NPS scores, reflecting overall improved candidate experience.

% In summary, quantitative and qualitative findings collectively suggest Zara effectively enhances the candidate experience by combining personalized, structured feedback with high-quality technical interactions and dynamic conversational flow, clearly validating the practical value of this implementation.

\section{Conclusion}

This paper presented Zara, a practical case study demonstrating how existing large language models (LLMs) can be effectively implemented to improve candidate support throughout the recruitment process. Zara is an AI-driven candidate support system developed at micro1, powered by OpenAI's GPT-4o, and designed to enhance several key recruitment stages. Specifically, Zara dynamically generates personalized practice interview scenarios and administers interactive, conversational AI-led interviews. More importantly, Zara provides structured and actionable post-interview feedback and autonomously resolves candidate queries—capabilities that significantly enhance the candidate experience. Delivering personalized feedback and addressing candidate questions at scale were previously impractical in traditional recruitment processes, highlighting the transformative impact of Zara’s implementation.

\subsection{Future Work}

Currently, Zara provides candidates with detailed feedback at their request. The next step in Zara’s evolution involves leveraging its analytical capabilities—specifically, insights gained from candidate interviews—to create personalized, adaptive skill-development courses. These individually tailored courses will incorporate targeted exercises, instructional content, and curated educational videos to directly address each candidate’s identified improvement areas. Ultimately, this proactive, personalized approach aims to significantly extend Zara’s candidate support capabilities, shifting from reactive feedback toward structured candidate skill development and career advancement.

\clearpage
\bibliographystyle{informs2014}
\bibliography{refs}

\clearpage

\end{document}